\documentclass[a4paper,10pt]{article}
\usepackage{amsmath}
\usepackage{amsthm}
\usepackage{amsfonts}
\usepackage{amscd}
\usepackage{amssymb}
\usepackage{MnSymbol}
\usepackage{wasysym}

\usepackage{amsfonts}
\usepackage{graphicx}
\usepackage{mathtools}
\usepackage{times}
\usepackage{bm}
\usepackage{natbib}
\usepackage{url}
\usepackage{thmtools}
\usepackage{fullpage}

\newcommand{\R}{\mathbb{R}}

\newcommand{\betastar}{\hat{\beta}^{\textstyle{*}}}
\newcommand{\betahat}{\hat{\beta}}
\newcommand{\betaw}{\hat{\beta}_w}
\newcommand{\betastarT}{\hat{\beta}^{\textstyle{*}^T}}
\newcommand{\bstar}{b^{\textstyle{*}}}
\newcommand{\bstarw}{\hat{\beta}^{\textstyle{*}}_w}

\newcommand{\wstar}{w^{\textstyle{*}}}

\newcommand{\epstar}{\varepsilon^{\textstyle{*}}}

\newcommand{\Ystar}{Y^{\textstyle{*}}}

\newcommand{\Sigstarhalf}{\widehat{\Sigma}^{\textstyle{*}^{1/2}}}
\newcommand{\Sigstarhalfinv}{\widehat{\Sigma}^{\textstyle{*}^{-1/2}}}
\newcommand{\lstar}{l^{\textstyle{*}}}

\newcommand{\Astar}{A^{\textstyle{*}}}
\newcommand{\Gostar}{\widehat{G}_o^{\textstyle{*}}}
\newcommand{\GostarT}{\widehat{G}_o^{\textstyle{*}^T}}
\newcommand{\Gohat}{\widehat{G}_o}
\newcommand{\GohatT}{\widehat{G}_o^T}
\newcommand{\Sigresstar}{\widehat{\Sigma}_{\text{res}}^{\textstyle{*}}}

\newcommand{\B}{\mathcal{B}}

\newcommand{\Sub}{\mathcal{S}}
\newcommand{\Q}{\mathcal{Q}}
\newcommand{\Proj}{\mathcal{P}}
\newcommand{\Env}{\mathcal{E}}
\newcommand{\Envspace}{\Env_{\Sigma}(\B)}

\newcommand{\dimEnv}{\text{dim}(\Env)}

\newcommand{\X}{\mathbb{X}}
\newcommand{\Y}{\mathbb{Y}}

\newcommand{\Sigres}{\widehat{\Sigma}_{\text{res}}}
\newcommand{\SigY}{\widehat{\Sigma}_{Y}}
\newcommand{\SigX}{\widehat{\Sigma}_{X}}

\newcommand{\betau}{\hat{\beta}_u}

\newcommand{\indep}{\rotatebox[origin=c]{90}{$\models$}}

\DeclareMathOperator{\Var}{var}

\newtheorem{thm}{Theorem}

\begin{document}





\title{Weighted envelope estimation to handle variability in model selection}

\author{Daniel J. Eck and R. Dennis Cook}


\maketitle

\begin{abstract}
Envelope methodology can provide substantial efficiency gains in multivariate 
statistical problems, but in some applications the estimation of the envelope 
dimension can induce selection volatility that may mitigate those gains. 
Current envelope methodology does not account for the added variance that can 
result from this selection. In this article, we circumvent dimension selection 
volatility through the development of a weighted envelope estimator. 
Theoretical justification is given for our estimator and validity of the 
residual bootstrap for estimating its asymptotic variance is established. A 
simulation study and an analysis on a real data set illustrate the utility of 
our weighted envelope estimator.
\end{abstract}

\noindent\emph{Keywords}: Dimension Reduction; Envelope Models; Model Selection; Residual Bootstrap; Variance Reduction.

\section{Introduction}

Envelope methodology was developed originally in the context of the 
multivariate linear regression model \citep*{cook},
\begin{equation}\label{mvmodel}
Y = \alpha + \beta X + \varepsilon,
\end{equation}
where $\alpha \in \R^r$, $\beta \in \R^{r\times p}$, the random response vector 
$Y \in \R^r$, the fixed predictor vector $X \in \R^p$ is centered to have mean 
zero, and the error vector $\varepsilon \sim N(0,\Sigma)$. Estimation is 
assumed to be based on $n$ independent samples from model \eqref{mvmodel} 
where $n > p$. It was shown by \citet*{cook} that the envelope estimator of 
the unknown coefficient matrix $\beta$ in \eqref{mvmodel} has the potential to 
yield massive efficiency gains relative to the maximum likelihood estimator of 
$\beta$. These efficiency gains can arise when the dimension $u$ of the 
envelope space, defined in the next section, is less than $r$. In most 
practical applications, $u$ is unknown and has to be estimated. This 
estimation can be problematic since the estimated variance of the envelope 
estimator is typically calculated conditional on the estimated dimension 
$\hat{u}$. Variation associated with model selection is therefore not 
considered in the current envelope paradigm.

In this article, we propose a weighted envelope estimator of $\beta$ that 
smooths out model selection volatility. The weighting is across all possible 
envelope models under \eqref{mvmodel}. The weights corresponding to each 
envelope estimator are functions of the Bayesian Information Criterion 
($\textsc{bic}$) value corresponding to that particular envelope model. 
Weighting in this manner is similar to the model averaging techniques 
discussed by \citet*{buckland} and \citet*{burnham} who provided a 
philosophical justification for the use of such weighted estimators without 
giving any theoretical properties. \citet{hjort} and 
\citet{liang} built on the philosophical justification for weighted estimators 
by deriving their asymptotic properties. \citet{claeskens} summarized 
extensions and applications of the theory of weighted estimators. However, 
these extensions do not include bootstrap techniques and do not encompass 
the framework of envelope models. Envelope models fit at dimensions greater 
than or equal to $u$ are all true non-nested data generating models and are 
ordered in preference from dimension $u$ to $r$. This context seems novel and 
is outside of the framework of \citet{claeskens}.

\section{The Envelope Model} 
The original motivation for envelope methodology came from the observation 
that, in the multivariate regression model \eqref{mvmodel}, some linear 
combinations of $Y$ may have a distribution that does not depend on $X$, while 
other linear combinations of $Y$ do depend on $X$. The envelope model 
separates out these immaterial and material parts of $Y$, and thereby allows 
for efficiency gains \citep{cook, su}.

More carefully, suppose that we can find a subspace $\Sub \subseteq \R^r$ so 
that
\begin{equation} \label{env-cond-1}
    \Q_{\Sub}Y \, \indep \, \Proj_{\Sub}Y \mid X, \quad \text{and} \quad
    \Q_{\Sub}Y \mid X = x_1 \sim \Q_{\Sub}Y \mid X = x_2, 
      \quad \text{for all} \quad x_1, x_2,
\end{equation}
where $\sim$ means identically distributed, $\Proj_{(\cdot)}$ projects onto 
the subspace indicated by its argument and $\Q = I_r - \Proj$. For any $\Sub$ 
with the properties \eqref{env-cond-1}, $\Proj_{\Sub}Y$ carries all of the 
material information and perhaps some of the immaterial information, while 
$\Q_{\Sub}$ contains just immaterial information. Let 
$\B = \text{span}(\beta)$ and $d = \text{dim}(\B)$ so that 
$0 < d \leq \min(p,r)$. Then \eqref{env-cond-1} holds if and only if 
$\B \subseteq \Sub$ and $\Sigma = \Sigma_{\Sub} + \Sigma_{\Sub^{\perp}}$, 
where $\Sigma_{\Sub} = \Var(\Proj_{\Sub}Y)$ and 
$\Sigma_{\Sub^{\perp}} = \Var(\Q_{\Sub}Y)$. The envelope is defined as the 
intersection of all subspaces $\Sub$ that satisfy \eqref{env-cond-1} and is 
denoted by $\Envspace$ with dimension $u = \text{dim}\{\Envspace\}$ satisfying 
$0 < d \leq u \leq r$.

The envelope model can be represented in terms of coordinates by 
parameterizing model \eqref{mvmodel} to incorporate conditions 
\eqref{env-cond-1}. Define $\Gamma \in \R^{r \times u}$ to be a 
semi-orthogonal basis matrix for $\Envspace$ and let 
 $(\Gamma,\Gamma_o) \in \R^{r \times r}$ be an orthogonal matrix. Then the 
envelope model with respect to model \eqref{mvmodel} is parameterized as 
\begin{equation} \label{envmodel}
  Y = \alpha + \Gamma\eta X + \varepsilon, \qquad \varepsilon \sim N(0,\Sigma),
\end{equation}
where $\Sigma = \Gamma\Omega\Gamma^T + \Gamma_o\Omega_o\Gamma_o^T$, 
$\Omega \in \R^{u \times u}$ and $\Omega_{o} \in \R^{(r-u) \times (r-u)}$ are 
positive definite, and $\eta \in \R^{u \times p}$ is $\beta = \Gamma\eta$ in 
the coordinates of $\Gamma$. We see from \eqref{envmodel}, that $\Envspace$ 
links the mean and covariance structures of the regression problem and it is 
this link that provides the efficiency gains. The gains can be massive when 
the immaterial information is large relative to the material information; for 
instance, when $\|\Omega\| \ll \|\Omega_o\|$, where $\|\cdot\|$ is a matrix 
norm \citep{cook}. An illuminating depiction and explanation of how 
an envelope increases efficiency in multivariate linear regression problems 
was given by \citet*[pgs. 134--135]{su}. \citet*{cook2} provided a more 
general framework for envelope methodology, which requires only a 
$\surd{n}$-consistent estimator $\hat{\theta}$ of an unknown parameter 
$\theta$ and a $\surd{n}$-consistent estimator of its asymptotic variability. 
\citet*{pls} showed that partial least squares gives a moment-based envelope 
estimator that is $\surd{n}$-consistent. As partial least squares is widely 
used in chemometrics and elsewhere, the \citet*{pls} finding indicates that 
envelope methodology is also widely applicable.

Candidate envelope estimators of $\beta$ at dimension $j$, denoted 
$\hat{\beta}_j$, are found via maximum likelihood estimation of model 
\eqref{envmodel} with $\hat{\beta}_j = \widehat{\Gamma}\hat{\eta}$. An 
estimator of $u$ is found by using a model selection criterion such as 
$\textsc{bic}$, Akaike Information Criterion ($\textsc{aic}$), likelihood 
ratio tests, or cross-validation. The estimated dimension $\hat{u}$ obtained 
from any one of these selection criteria is a variable quantity dependent on 
the observed data. Current envelope methodology does not address this extra 
variability. In the next two sections, we develop properties of a weighted 
estimator that takes this extra variability into account.

\section{ $\textsc{bic}$  Weighted Estimators}


The weighted estimator that we consider is of the form
\begin{equation} \label{wt-beta}
  \hat{\beta}_w = \sum_{j=1}^r w_j\hat{\beta}_{j},
\end{equation}  
where $\sum_{j=1}^r w_j = 1$ and $w_j \geq 0$, for $j = 1,...,r$. The weights 
$w_j$ depend on the $\textsc{bic}$ values for all of the candidate envelope 
models under consideration. Let the $\textsc{bic}$ value for the envelope 
model with dimension $j$ be denoted by 
$b_j = -2l(\hat{\beta}_j) + k(j)\log(n)$, where $l(\hat{\beta}_j)$ is the log 
likelihood evaluated at the envelope estimator $\hat{\beta}_j$ and 
$
  k(j) = r + pj + r(r+1)/2
$ 
is the number of parameters of the envelope model of dimension $j$. The 
weight for envelope model $j$ is constructed as
\begin{equation} \label{weights}
  w_j =  \frac{\exp(-b_j)}{\sum_{k=1}^r \exp(-b_k)}.
\end{equation}
It follows from arguments in the Supplement that $\betaw$ is a 
$\surd{n}$-consistent estimator of $\beta$, but assessing the variance of 
$\betaw$ is not so straightforward. In the next section, we show that the 
residual bootstrap provides a consistent estimator of $\Var(\betahat_u)$. We 
use $\textsc{bic}$ in \eqref{weights} because, in ours and others' 
experiences, $\textsc{bic}$ performs well when selecting the dimension of an 
envelope model. $\textsc{aic}$ tends to overselect the true dimension of an 
envelope model, likelihood ratio testing is inconsistent, and cross-validation 
is primarily used in prediction problems. We do not claim that $\textsc{bic}$ 
is optimal in this application.

\section{Bootstrap for $\hat{\beta}_w$}

The envelope estimator $\hat{\beta}_u$ at the true dimension $u$ is 
$\surd{n}$-consistent and asymptotically normal \citep*{cook, cook2}. The 
residual bootstrap used to estimate the variability of $\hat{\beta}_u$ uses 
the starred responses,
\begin{equation} \label{env-boot}
  \Ystar = \X\hat{\beta}_u^T + \epstar,
\end{equation}
to obtain $\betastar_u$, where $\X \in \R^{n\times p}$ is the fixed design 
matrix with rows $X_i^T$ and the rows of $\epstar \in \R^{n\times r}$ are the 
realizations of $n$ resamples of the residuals from the ordinary least squares 
fit of \eqref{mvmodel}. This process is performed a total of $B$ times with a 
new $\betastar_u$ computed from \eqref{env-boot} at each iteration. The setup 
in \citet*[Section 2, pgs. 122-124 and Theorem 2]{andrews} confirms that the 
sample variance of the $\betastar_u$s provides a $\surd{n}$-consistent 
estimator of the asymptotic variability of $\hat{\beta}_u$. 
The problem with this approach, as it currently stands, is that $u$ is unknown. 
The current implementation of the residual bootstrap implicitly assumes that 
$\hat{u} = u$. Therefore, variability introduced by model selection 
uncertainty is ignored. This issue is resolved by using $\hat{\beta}_w$ in 
place of $\hat{\beta}_u$ in \eqref{env-boot}. The next theorem formalizes our 
asymptotic justification for the use of the weighted envelope estimator 
$\hat{\beta}_w$ in practical problems. Its proof is given in the Supplement.

\begin{thm} \label{weight-boot}
Assume regression model \eqref{mvmodel} and suppose that an envelope subspace 
of dimension $u = 1,...,r$ exists. Assume that 
$\SigX = n^{-1}\X^T\X \to \Sigma_{X} > 0$. Let $\hat{\beta}_w$ be the weighted 
envelope estimator of $\beta$ defined in \eqref{wt-beta} and let 
$\hat{\beta}^{\textstyle{*}}_w$ be the weighted envelope estimator of $\beta$ 
obtained from resampled data. Then, as $n$ tends to $\infty$, 
\begin{equation} \label{opmn}
  \begin{split}
    &\surd{n}\left\{ \text{vec}(\betastar_{w}) 
      - \text{vec}(\hat{\beta}_{w}) \right\} 
    = \surd{n}\left\{ \text{vec}(\betastar_{u}) 
      - \text{vec}(\hat{\beta}_{u}) \right\} \\
    &\qquad+ O_p\left\{n^{(1/2-p)}\right\} 
      + 2(u-1)O_p(1)\surd{n}e^{-n \mid O_p(1)\mid}. 
  \end{split}
\end{equation}
\end{thm}

Theorem \ref{weight-boot} shows the utility of the weighted envelope estimator 
$\betaw$. In \eqref{opmn}, we see that the asymptotic distribution of the 
residual bootstrap at $\betaw$ is the same as the asymptotic distribution of 
the residual bootstrap at $\betau$. The difference between the two bootstrap 
procedures is that the bootstrap given in Theorem~\ref{weight-boot} does not 
require the conditioning on $\hat{u}$ as a prerequisite for its 
implementation. 

The orders in \eqref{opmn} result from model selection variability that arises 
from four sources. The 
$
  O_p\left\{n^{(1/2-p)}\right\}
$
term corresponds to the rate at which $\surd{n}w_j$ and $\surd{n}\wstar_j$ 
vanish for $j = u+1,...r$. This rate is a cost of over estimation of the 
envelope space. It decreases quite fast, particularly when $p$ is not small, 
because models with $j > u$ are true and thus have no systematic bias due to 
choosing the wrong dimension. 

The 
$
  2(u-1)\surd{n}e^{-n\mid O_p(1)\mid} 
$
term corresponds to the rate at which $\surd{n}w_j$ and $\surd{n}\wstar_j$ 
vanish for $j = 1,...,u-1$. This rate arises from under estimating the 
envelope space and it is affected by systematic bias arising from choosing the 
wrong dimension. To gain intuition about this rate,  let 
$
  B_{j} = \left(G_o^T\Sigma G_o\right)^{-1/2}G_o^T\beta\Sigma_X^{1/2},
$
where $G_o \in \R^{r \times (r-j)}$ is the population basis matrix for the 
complement of the envelope space of dimension $j$.  This quantity is a 
standardized version of $G_o^T\beta$ that reflects bias, since 
$G_o^T\beta \neq 0$ when $j < u$, but $G_o^T\beta = 0$ when $j \geq u$. Let 
$\widehat{B}_{j,n}$ denote the $\surd{n}$-consistent estimator of $B_{j}$ 
obtained by plugging in the sample version of $\Sigma_{X}$ and the estimators 
of $G_{o}$, $\Sigma$ and $\beta$ that arise by maximizing the likelihood with 
dimension $j < u$. Then the $-n\mid O_p(1)\mid$ term appearing in the exponent 
of $2(u-1)\surd{n}e^{-n\mid O_p(1)\mid}$ is the rate at which 
$-n\log(\mid I_{p}+\widehat{B}_{j,n}^{T}\widehat{B}_{j,n}\mid)$ approaches $-\infty$. 
Additionally, this term is 0 when $u=1$. That arises because we consider only 
regressions in which $\beta \neq 0$ and thus $u \geq 1$. When $u=1$ 
under estimation is not possible in our context and thus 
$2(u-1)\surd{n}e^{-n\mid O_p(1)\mid}$ vanishes.

The weights in \eqref{weights} differ from those 
mentioned in \citet*{burnham} which were also advocated by \citet*{kass} and 
\citet*{tsague}. These weights are of the form 
\begin{equation} \label{weights-kass}
  \tilde{w}_j = \frac{\exp(-b_j/2)}{\sum_{k=1}^r\exp(-b_k/2)}
\end{equation}  
and they correspond to an approximation of the posterior probability for model 
$j$ given the observed data under the prior that places equal weight for all 
candidate models. Weights of the form \eqref{weights-kass} do not have the 
same asymptotic properties as the weights given by \eqref{weights}. When 
$p=1$, the term $\surd{n}\tilde{w}_{j=u+1}$ defined by \eqref{weights-kass} 
does not vanish as $n\to\infty$. We therefore would not have the same 
asymptotic result given by \eqref{opmn} in Theorem~\ref{weight-boot}. Instead, 
there would be non-zero weight placed on the envelope model with dimension 
$j=u+1$ asymptotically. This weighting scheme would therefore lead to higher 
estimated variability than is necessary in practice. However, this issue is no 
longer problematic when $p > 1$. When $p > 1$ and weights 
\eqref{weights-kass} are used, the $O_p\left\{n^{(1/2-p)}\right\}$ term in 
\eqref{opmn} becomes $O_p\left\{n^{(1-p)/2}\right\}$, resulting in a slower 
rate of convergence.

Constructing $\hat{\beta}_w$ with respect to $\textsc{bic}$ may not be 
the only weighting scheme that satisfies 
\begin{equation} \label{opmn2}
  \surd{n}\left\{ \text{vec}(\betastar_{w}) 
    - \text{vec}(\hat{\beta}_{w}) \right\} 
  = \surd{n}\left\{ \text{vec}(\betastar_{u}) 
    - \text{vec}(\hat{\beta}_{u}) \right\}
    + O_p\left\{f(p,n)\right\} 
\end{equation}
where $f(p,n)$ is a function that depends on how the weights are constructed. 
Any weighting scheme such that, for all $j \neq u$,
\begin{equation} \label{opmn-cond}
\surd{n}\left\{ \text{vec}(\betastar_{j}) 
  - \text{vec}(\hat{\beta}_{j}) \right\} \to 0
\end{equation}
as $n\to\infty$ satisfies \eqref{opmn2}. Weighting schemes that violate 
\eqref{opmn-cond} will not result in a bootstrap that is consistent.

Similar weights with $\textsc{aic}$ in place of $\textsc{bic}$ do not satisfy 
\eqref{opmn-cond}. Interchanging $\textsc{bic}$ with $\textsc{aic}$ in the 
proof of Theorem~\ref{weight-boot} produces weights of the form 
$w_j = \mid O_p(1)\mid e^{2\{k(u)-k(j)\}}$ for all $j = u+1,...,r$ which do not vanish as 
$n\to\infty$.


\section{Examples}
We now provide three examples which show the utility of 
Theorem~\ref{weight-boot}. The first two are simulated examples in which we 
know $\beta$, $\Sigma$, $u$, and $\Proj_{\Envspace}$. The third is based on real 
data.

\subsection{Simulated examples}

\textit{Example 1:} For this example we create a setting in which 
$Y \in \R^3$ is generated according to the model
\begin{equation} \label{simmodel}
  Y_i = \beta X_i + \varepsilon_i, \qquad \varepsilon_i 
    \overset{ind}{\sim} N(0,\Sigma),
\end{equation}
$(i = 1,...,n)$, where $X_i \in \R^2$ is a continuous predictor with entries 
generated independently from a normal distribution with mean 4 and variance 1. 
The covariance matrix $\Sigma$ was generated using three orthonormal vectors 
and has eigenvalues of $50$, $10$, and $0.01$. The matrix 
$\beta \in \R^{3 \times 2}$ is an element in the space spanned by the second 
and third eigenvectors of $\Sigma$. We know that the dimension of 
$\mathcal{E}_{\Sigma}(\B)$ is $u = 2$.

\begin{table}[h!]
\begin{center}
\begin{tabular}{lcccc}
                                                           & $n = 50$ & $n = 100$ &  $n = 500$   & $n = 2000$ \\
$ \|\text{vec}(\betaw)-\text{vec}(\hat{\beta}_{u=2})\|_2$  & 2.3      & 0.016     &  $\approx 0$ & $\approx 0$ \\
$ \|\widehat{\Var}(\bstarw-\hat{\beta}_{u=2})\| $          & 0.18     & 0.12      &  0.021       & 0.0051 
\end{tabular}
\end{center}
\caption{ Comparison of $\betaw$ and $\hat{\beta}_{u=2}$. 
  The first row is the Euclidean difference between $\text{vec}(\betaw)$ and 
  $\text{vec}(\hat{\beta}_{u=2})$ from the original dataset. The second row 
  is the spectral norm of the estimated variance of the difference of all 
  bootstrap realizations of $\bstarw$ and $\hat{\beta}_{u=2}$ with bootstrap 
  sample size $B = n$.}
\label{env-ex1-tab1}
\end{table}

Four datasets were simulated under  
model \eqref{simmodel} at different sample sizes. The multivariate residual 
bootstrap was used to compare the weighted envelope estimator $\betaw$ with the 
oracle envelope estimator $\hat{\beta}_{u=2}$ across the simulated datasets. 
In Table~\ref{env-ex1-tab1}, we see that the Euclidean difference of 
$\text{vec}(\hat{\beta}_{u=2})$ and $\text{vec}(\betaw)$ shrinks as $n$ 
increases, and that the spectral norm of the variance of differences also 
shrinks as $n$ increases. Taken together, these findings support the 
conclusions of Theorem~\ref{weight-boot}.

\textit{Example 2:} For this example we illustrate the effect that $p$ has on 
the performance of the weighted envelope estimator. We generated data 
according to model \eqref{simmodel} with $Y \in \R^5$. In this example $u=1$ 
and $\Sigma$ is compound symmetric with diagonal entries set to 1 and 
off-diagonal entries set to 0.5, $\beta = 1_{r}c_{p}^{T}$, where $1_{r}$ is 
the $r \times 1$ vector of ones, and $c_{p}$ is a $p \times 1$ vector where 
every entry is 10. We generate the predictors according to $X \sim N(0, I_p)$, 
where $I_p$ is the $p$-dimensional identity matrix. We set $n = 250$.

We then perform a residual bootstrap with sample size $B = 250$ 
and, for each $p$ considered, we report the number of times each dimension was 
selected by $\textsc{bic}$, denoted by $n(\hat{u})$. From 
Table~\ref{env-ex2-tab1}, we see that the distribution of $\hat{u}$, across the 
$B$ resamples, approaches a point mass at the truth as $p$ increases with $u$ 
fixed. This implies that our bootstrap procedure improves as 
$p$ increases with $u$ fixed, as indicated by Theorem~\ref{weight-boot}.

\begin{table}[h!]
\begin{center}
\begin{tabular}{lccc}
  & $n(\hat{u} = 1)$ & $n(\hat{u} = 2)$ & $n(\hat{u} = 3)$  \\
  $p = 2$  & 128 & 111 & 11 \\
  $p = 5$  & 214 &  34 &  2 \\
  $p = 10$ & 249 &   1 &  0 \\
  $p = 25$ & 250 &   0 &  0 \\
\end{tabular}
\end{center}
\caption{ The bootstrap distribution of $\hat{u}$ as $p$ increases, where 
  $\hat{u}$ is selected by $\textsc{bic}$ and $n(\hat{u} = j)$ is the number 
  of times \textsc{bic} selected envelope dimension $j$.} 
\label{env-ex2-tab1}
\end{table}

\subsection{Cattle data}

The data in this example, analyzed in \citet*{kenward} and \citet*{cook2}, 
came from an experiment that compared two treatments for the control of a 
parasite in cattle. The experimenters were interested in finding if the 
treatments had differential effects on weight and, if so, about when they 
first occurred. There were sixty animals in this experiment and thirty 
animals were randomly assigned to the two treatments. Their weights (in 
kilograms) were then recorded at weeks 2, 4,..., 18 and 19 after treatment 
\citep*{kenward}. In our analysis, we considered the multivariate linear model 
\eqref{mvmodel}, where $Y_i \in \R^{10}$ is the vector of cattle weights from 
week 2 to week 19, and predictor $X_i$ is either 0 or 1 indicating which of 
the two treatments was assigned. In this model, $\alpha$ is the mean profile 
for one treatment and $\beta$ is the mean difference between the two 
treatments. 

Since the two treatments were not expected to have an immediate 
measurable affect on weight, some linear combinations of the response vector 
are not expected to depend on the treatment. Therefore the envelope model 
\eqref{envmodel} is expected to perform well in this application because of 
our belief that \eqref{env-cond-1} holds with $\Env_{\Sigma}^{\perp}(\B)$ at 
least as large as the span of the linear combinations that isolate the first 
few elements of the response vector.

Envelope models were fitted at each dimension from $1$ to $10$. The 
likelihood ratio test selected $\hat{u} = 1$ and $\textsc{bic}$ selected 
$\hat{u} = 3$ as the dimension of the envelope model. Further complicating 
matters, when $\textsc{bic}$ is used to determine $u$ at every resample of the 
multivariate residual bootstrap with sample size $B = 60$, we see 
high variability in the models selected. Specifically, 
$n(\hat{u} = 1) = 10$, $n(\hat{u} = 2) = 10$, $n(\hat{u} = 3) = 24$, 
$n(\hat{u} = 4) = 12$, and $n(\hat{u} = 5) = 4$. Model selection variability  
of this variety is precisely the reason why the weighted envelope estimator is 
advocated. 

In Table~\ref{env-ex2-tab3}, we see the ratios of bootstrapped estimated 
standard errors for envelope estimators to those of the maximum likelihood 
estimator of the $\beta$ from the full model \eqref{mvmodel}, 
$\text{se}^{\textstyle{*}}(\betahat_r)/\text{se}^{\textstyle{*}}(\betahat_w)$, 
averaged across 25 replications. Standard errors of the averaged ratios across 
replications are all less than 7\% of the reported ratios and 
the average standard error is 2.6\% of the reported ratio.  
Ratios greater than 1 indicate that the envelope 
estimator is more efficient than the standard estimator. We see that 
$\hat{\beta}_w$ is comparable to $\hat{\beta}_{u=3}$. Similar conclusions are 
drawn from the other elements of estimates of $\beta$. The findings displayed 
in Table~\ref{env-ex2-tab3} illustrate that the weighted envelope estimator 
can provide useful efficiency gains while properly accounting for model 
selection variability.

\begin{table}[h!]
\begin{center}
\begin{tabular}{ccccccc}
$B$ & $\hat{\beta}_w$ & $\hat{\beta}_{u=1}$ & $\hat{\beta}_{u=2}$ & 
  $\hat{\beta}_{u=3}$ & $\hat{\beta}_{u=4}$ & $\hat{\beta}_{u=5}$ \\
  60   &  1.98  &  5.54  &  3.05  &  1.69  &  1.31  &  1.23  \\
  100  &  1.97  &  5.54  &  2.55  &  1.54  &  1.32  &  1.21  \\
  500  &  1.82  &  5.47  &  2.78  &  1.57  &  1.31  &  1.16  \\
  2000 &  1.81  &  5.37  &  2.60  &  1.53  &  1.29  &  1.16  \\
\end{tabular}
\end{center}
\caption{Averaged ratios of estimated standard errors across 25 replications 
  of the multivariate residual bootstrap at different numbers of resamples $B$ 
  for the fifth element of estimates of $\beta$. Standard errors of the 
  averaged ratios are in parentheses. } 
\label{env-ex2-tab3}
\end{table}

We next report results of a simulation study using the cattle data 
to show further support for Theorem~\ref{weight-boot}. We generate data 
according to the model 
$$
  Y_i = \alpha + \beta X_i + \varepsilon_i, 
    \qquad \varepsilon_i \overset{ind}{\sim} 
    N(0,\Sigma),
$$
$(i = 1,...,n)$ where $\alpha$, $\beta$, and $\Sigma$ were set to the 
estimates obtained from the envelope model fit to the cattle data at dimension 
$u = 3$, 
and $X_i$ is the binary indicator that specified treatment. Cows are split 
evenly between the two treatment groups and the assignment was random.

In Table~\ref{env-ex2-tab4}, we see that the Euclidean 
differences between $\text{vec}(\hat{\beta}_{u=3})$ and $\text{vec}(\betaw)$ 
shrink as $n$ increases. The same is true for the differences 
between $\text{vec}(\hat{\beta}_{u=4})$ and $\text{vec}(\betaw)$. This was 
expected since the envelope model fit with $u=4$ is a true data generating 
model. However, we see that the Euclidean distance between 
$\text{vec}(\hat{\beta}_{u=2})$ and $\text{vec}(\betaw)$ does not shrink as 
$n$ increases. Again, this was expected since the envelope model fit with $u=2$ 
is not a true data generating model. These simulation results are in alignment 
with the conclusions of Theorem~\ref{weight-boot}.

\begin{table}[h!]
\begin{center}
\begin{tabular}{lcccc}
                                                           & $n = 60$ & $n = 100$ & $n = 500$  & $n = 2000$ \\
$ \|\text{vec}(\betaw)-\text{vec}(\hat{\beta}_{u=2})\|_2$  & 9.36     & 0.83      & 0.91       & 4.2        \\
$ \|\text{vec}(\betaw)-\text{vec}(\hat{\beta}_{u=3})\|_2$  & 9.37     & 0.54      & 0.070      & 0.00028    \\
$ \|\text{vec}(\betaw)-\text{vec}(\hat{\beta}_{u=4})\|_2$  & 9.37     & 0.69      & 0.34       & 0.090      \\
\end{tabular}
\end{center}
\caption{ Comparison of $\betaw$ and $\hat{\beta}_{u=2}$, 
  $\hat{\beta}_{u=3}$, and $\hat{\beta}_{u=4}$. 
  The rows are the Euclidean difference between $\text{vec}(\betaw)$ and 
  the indicated envelope estimator from the original dataset. }
\label{env-ex2-tab4}
\end{table}

\section{Discussion}

\citet*{efron} proposed an estimator motivated by bagging \citep*{breimen} 
that aims to reduce variability and smooth out discontinuities resulting from 
model selection volatility. Variability of the model averaged estimator of 
\citet*{efron} is assessed via a double bootstrap. These techniques have been 
applied to envelope methodology in \citet{eck} and useful variance reduction 
was found empirically. The problem of interest in \citet{eck} falls outside 
the scope of the multivariate linear regression model, and general envelope 
methodology \citep*{cook2} was required to obtain efficiency gains. 
n the context of the multivariate linear regression model, we 
showed that only a single level of bootstrapping is necessary.

The idea of weighting envelope estimators across all candidate 
dimensions extends to partial least squares \citep{pls}, predictor 
envelopes \citep{predictor}, and sparse response envelopes \citep{sparse}.

\section{Supplementary material}
Supplementary material available at Biometrika online includes the proof of 
Theorem~\ref{weight-boot} and a complete version of Table~\ref{env-ex2-tab3} 
that includes standard errors for all of the averaged ratios.

\newpage

\section*{`Supplementary material for Weighted envelope estimation to handle variability in model selection'}

This Supplementary Materials section contains the proof of 
Theorem~\ref{weight-boot} and an extended version of Table~\ref{env-ex2-tab3} 
in \citet*{eck2}. 

In Table~\ref{env-ex2-tab5}, we see the ratios of bootstrapped estimated 
standard errors between envelope estimators to those of the maximum likelihood 
estimator of the $\beta$ from the full model \eqref{mvmodel}, 
$\text{se}^{\textstyle{*}}(\betahat_r)/\text{se}^{\textstyle{*}}(\betahat_w)$, 
averaged across 25 replications. Standard errors of the averaged ratios across 
replications are in parentheses.

\begin{table}[h!]
\begin{center}
\begin{tabular}{ccccccc}
$B$ & $\hat{\beta}_w$ & $\hat{\beta}_{u=1}$ & $\hat{\beta}_{u=2}$ & 
  $\hat{\beta}_{u=3}$ & $\hat{\beta}_{u=4}$ & $\hat{\beta}_{u=5}$ \\
  60   &  1.98 (0.081)  &  5.54 (0.14)   &  3.05 (0.19)   &  1.69 (0.11)   &  1.31 (0.044)   &  1.23 (0.039)  \\
  100  &  1.97 (0.10)   &  5.54 (0.14)   &  2.55 (0.15)   &  1.54 (0.044)  &  1.32 (0.038)   &  1.21 (0.027)  \\
  500  &  1.82 (0.031)  &  5.47 (0.074)  &  2.78 (0.076)  &  1.57 (0.024)  &  1.31 (0.013)   &  1.16 (0.013)  \\
  2000 &  1.81 (0.017)  &  5.37 (0.049)  &  2.60 (0.032)  &  1.53 (0.013)  &  1.29 (0.0084)  &  1.16 (0.0071) \\
\end{tabular}
\end{center}
\caption{Averaged ratios of estimated standard errors across 25 replications 
  of the multivariate residual bootstrap at different numbers of resamples $B$ 
  for the fifth element of estimates of $\beta$. Standard errors of the 
  averaged ratios are in parentheses. } 
\label{env-ex2-tab5}
\end{table}

Here is the proof of Theorem~\ref{weight-boot} in \citet*{eck2}:

\begin{proof} 
We go through the steps showing that \eqref{opmn} in \citet*{eck2} holds. Recall that $u = \dimEnv$. Define $l(\hat{\beta}_j)$ to be the log likelihood of the envelope model evaluated at the envelope estimator $\hat{\beta}_j$, fitting with $\dimEnv = j$, and define $k(j)$ to be the number of parameters of the envelope model of dimension $j$. From the construction of $b_j$ and the above calculations we see that
$$
  e^{b_u-b_j} = e^{-2\{l(\hat{\beta}_u) - l(\hat{\beta}_j)\}}n^{-\{k(j) - k(u)\}}.
$$
Let $\bstar_j$ be the \textsc{bic} value of the envelope model of dimension $j$ fit to the starred data and define
$$
  \wstar_j = \frac{e^{-\bstar_j}}{\sum_{k=1}^r e^{-\bstar_k}}.
$$
Let $\|\cdot\|$ be the Euclidean norm. We show that 
$
  \surd{n}\left\{\wstar_j\text{vec}(\betastar_j) - w_j\text{vec}(\hat{\beta}_j)\right\} \to 0 
$ 
for $j \neq u$ by showing that 
$$
  \surd{n}\|\wstar_j\text{vec}(\betastar_j) - w_j\text{vec}(\hat{\beta}_j)\| 
    \; \leq \; \surd{n}\|\wstar_j\text{vec}(\betastar_j)\| + \surd{n}\|w_j\text{vec}(\hat{\beta}_j)\| \; \to \; 0
$$
as $n\to\infty$ for all $j \neq u$. Now,
\begin{equation} \label{weightineq-1}
  \begin{split}
    &\surd{n}w_j\|\text{vec}(\hat{\beta}_j)\| \leq \surd{n}\mid O_p(1)\mid e^{b_u-b_j} \\
      &\qquad= \mid O_p(1) \mid n^{\left\{k(u)-k(j)+1/2\right\}}e^{ -2\left\{l(\hat{\beta}_u) - l(\hat{\beta}_j)\right\}} \\
      &\qquad= \mid O_p(1) \mid n^{\left\{k(u)-k(j)+1/2\right\}}e^{ 2\left\{l(\hat{\beta}_r) - l(\hat{\beta}_u)\right\} - 2\left\{l(\hat{\beta}_r) - l(\hat{\beta}_j)\right\}}.
  \end{split}  
\end{equation}
The first inequality in \eqref{weightineq-1} follows from the fact that $\|\text{vec}(\hat{\beta}_j)\| \leq \|\text{vec}(\hat{\beta}_r)\|$ and $\|\text{vec}(\hat{\beta}_r)\| = O_p(1)$. We first consider the case where $j = u+1,...,r$. In this setting, models with envelope dimensions $u$ and $j$ are both true and nested within the full model with envelope dimension $r$. Consequently, $-2\{l(\hat{\beta}_u) - l(\hat{\beta}_r)\}$ and $-2\{l(\hat{\beta}_j) - l(\hat{\beta}_r)\}$ are asymptotically distributed as $\chi^2_{p(r-u)}$ and $\chi^2_{p(r-j)}$ by Wilks' Theorem. Therefore $e^{-2\{l(\hat{\beta}_u) - l(\hat{\beta}_j)\}} = O_{p}(1)$ since it is the exponentiation of the difference between two $\chi^{2}$ random variables. We see that
$$
  \surd{n}w_j\|\text{vec}(\hat{\beta}_j)\| \leq  \mid O_p(1)\mid  n^{\left\{k(u)-k(j)+1/2\right\}} = O_p\left[n^{\left\{k(u)-k(j)+1/2\right\}}\right].
$$
Since $j > u$, we have that $k(u)-k(j) = p(u-j) \leq -p$. Thus, 
$$
  \surd{n}w_j\|\text{vec}(\hat{\beta}_j)\| \leq O_p\left\{n^{\left(1/2 - p\right)}\right\}
$$
for $j = u+1,...,r$. Following the same steps as \eqref{weightineq-1}, applied to the starred data, yields
\begin{equation} \label{weightineq-1-star}
  \surd{n}\wstar_j\|\text{vec}(\betastar_j)\| \leq \mid O_p(1)\mid n^{\left\{k(u)-k(j)+1/2\right\}}e^{ -2\left\{\lstar(\betastar_u) - \lstar(\betastar_r)\right\} + 2\left\{\lstar(\betastar_j) - \lstar(\betastar_r)\right\} }
\end{equation}
where $\lstar(\cdot)$ is the log likelihood function corresponding to the starred data. Both $-2\left\{\lstar(\betastar_u) - \lstar(\betastar_r)\right\}$ and $2\left\{\lstar(\betastar_j) - \lstar(\betastar_r)\right\}$ in \eqref{weightineq-1-star} are $O_p(1)$. Thus, 
$$
  \surd{n}w_j\|\text{vec}(\betastar_j)\| \leq \; \mid O_p(1)\mid  n^{\left\{k(u)-k(j)+1/2\right\}} = O_p\left[n^{\left\{k(u)-k(j)+1/2\right\}}\right],
$$
and,
$
  \surd{n}w_j\|\text{vec}(\betastar_j)\| \leq O_p\left\{n^{\left(1/2 - p\right)}\right\}
$
for all $j = u+1,...,r$. This establishes that 
$$
  \surd{n}\|\wstar_j\text{vec}(\betastar_j) - w_j\text{vec}(\hat{\beta}_j)\| \leq O_p\left\{n^{\left(1/2-p\right)}\right\},
$$
for $j = u+1,...,r$.

Turning to the case when $j = 1,...,u-1$, consider the exponent $e^{-\lambda_j}$, with $\lambda_j = 2\left\{l(\hat{\beta}_r) - l(\hat{\beta}_j)\right\}$. This is a log likelihood ratio although, unlike the case when $j = u+1,...,r$, it does not follow a $\chi^{2}$ distribution asymptotically.  Let $\widehat{G}$ and $\widehat{G}_o$ be the estimated bases for the envelope space and its orthogonal complement fitting with dimension $j = 1,...,u-1$, so $\widehat{G} \in \R^{r \times j}$ and $\widehat{G}_{o} \in \R^{r \times (r-j)}$. We write
\begin{align} 
    \lambda_j = & \; 2\left\{l(\hat{\beta}_r) - l(\hat{\beta}_j)\right\} \nonumber \\
      = & \; n\log\mid\widehat{G}^T\Sigres\widehat{G}\mid + n\log\mid\widehat{G}_o^T\SigY\widehat{G}_o\mid - n\log\mid\Sigres\mid \nonumber \\
      = & \; n\log\mid\widehat{G}^T\Sigres\widehat{G}\mid + n\log\mid\widehat{G}_o^T\Sigres\widehat{G}_o\mid - n\log\mid\Sigres\mid \nonumber \\
        &\qquad+ n\log\mid I_{p} + \widehat{\Sigma}_X^{1/2}\hat{\beta}_r^T\widehat{G}_{o}\left(\widehat{G}_{o}^T\Sigres\widehat{G}_{o}\right)^{-1}\widehat{G}_{o}^T\hat{\beta}_r\widehat{\Sigma}_X^{1/2}\mid \label{lambdaj1}
\end{align}
where $\SigY = n^{-1}\Y^T\Y$. The second equation in \eqref{lambdaj1} follows by applying the usual expansion of the determinant of a sum of the form $A + BB^T$. To see this,
\begin{align*}
  \mid \GohatT\SigY\Gohat \mid &= \mid \GohatT\Sigres\Gohat + \GohatT\Y^T\X(\X^T\X)^{-1}\X^T\Y\Gohat \mid \\
    &= \mid \GohatT\Sigres\Gohat + \GohatT\betahat_r\SigX\betahat_r^T\Gohat \mid \\
    &= \mid \GohatT\Sigres\Gohat \mid \times \mid I_p + \SigX^{1/2}\betahat^T_r\Gohat\left(\GohatT\Sigres\Gohat\right)^{-1}\GohatT\betahat_r\SigX^{1/2}\mid,
\end{align*}
where $\GohatT\betahat_r\SigX\betahat_r^T\Gohat = \GohatT\Y^T\X(\X^T\X)^{-1}\X^T\Y\Gohat$ because of the definition of $\betahat_r = \Y^T\X(\X^T\X)^{-1}$.

We bound $\lambda_j$ from below by further minimizing the first three addends in \eqref{lambdaj1} over $(\widehat{G},\widehat{G}_o$). These are minimized globally when the columns of $\widehat{G}$ span any reducing subspace of $\Sigres$ and is $0$ at the minimum. Thus
\begin{equation} \label{Ajn}
  \begin{split}
    \lambda_j &\geq n\log\mid I_{p} + \widehat{\Sigma}_X^{1/2}\hat{\beta}_r^T\widehat{G}_{o}\left(\widehat{G}_{o}^T\Sigres\widehat{G}_{o}\right)^{-1}\widehat{G}_{o}^T\hat{\beta}_r\widehat{\Sigma}_X^{1/2}\mid \\
    &= n\log\mid I_{p} + \widehat{\Sigma}_X^{1/2}\hat{\beta}_r^T\Sigres^{-1/2}\left\{\Sigres^{1/2}\widehat{G}_{o}\left(\widehat{G}_{o}^T\Sigres\widehat{G}_{o}\right)^{-1}\widehat{G}_{o}^T\Sigres^{1/2}\right\}\Sigres^{-1/2}\hat{\beta}_r\widehat{\Sigma}_X^{1/2}\mid \\
    &= n\log(\widehat{A}_{j,n}),
  \end{split}
\end{equation}
where $\widehat{A}_{j,n}$ is defined implicitly. The quantity 
$
  \Sigres^{1/2}\widehat{G}_{o}\left(\widehat{G}_{o}^T\Sigres\widehat{G}_{o}\right)^{-1}\widehat{G}_{o}^T\Sigres^{1/2}
$
in \eqref{Ajn} is the projection into the column space of $\Sigres^{1/2}\widehat{G}_{o}$. The quantity $\widehat{G}_{o}^T\betahat_r \neq 0$ almost surely since $j=1,...,u-1$. As a result, the column space of $\Sigres^{-1/2}\hat{\beta}_r\widehat{\Sigma}_X^{1/2}$ in \eqref{Ajn} has a nontrivial intersection with the column space of $\Sigres^{1/2}\widehat{G}_{o}$ almost surely. Therefore $\widehat{A}_{j,n} > 1$ almost surely. We can write $n\log(\widehat{A}_{j,n}) = n\mid O_p(1) \mid$ and we have the bound
$$
  e^{-\lambda_j} = e^{-2\{l(\hat{\beta}_j) - l(\hat{\beta}_r)\}} \leq e^{-n\log(\widehat{A}_{j,n})} = e^{-n\mid O_p(1)\mid}.
$$
Therefore,
\begin{equation} \label{weightineq-2}
  \begin{split}
    &\log(w_{j}) \leq b_u - b_j \\
      &\qquad =  -2\{l(\hat{\beta}_u) - l(\hat{\beta}_r)\} + 2\{l(\hat{\beta}_j) - l(\hat{\beta}_r)\} + \{k(u)-k(j)\}\log(n) \\
      &\qquad =  |O_{p}(1)| - \lambda_{j} + \{k(u)-k(j)\}\log(n) \\
      &\qquad \leq |O_{p}(1)|  - n\mid O_p(1)\mid + \{k(u)-k(j)\}\log(n) = -n\mid O_p(1)\mid
  \end{split}
\end{equation}
and we see that $\surd{n}w_j \leq \surd{n}e^{-n\mid O_p(1)\mid}$ for $j = 1,...,u-1$. 

Define $\Gostar$ to be the estimate of $G_o$ obtained from the starred data and let 
\begin{equation} \label{Ajnstar}
  \begin{split}
    \Astar_{j,n} &= \mid I_p + \widehat{\Sigma}_X^{1/2} \betastarT_r \Gostar\left(\GostarT \Sigresstar\Gostar\right)^{-1}\GostarT\betastar_r\widehat{\Sigma}_X^{1/2}\mid \\
      &= \mid I_p + \widehat{\Sigma}_X^{1/2} \betastarT_r\Sigstarhalfinv \left\{\Sigstarhalf\Gostar\left(\GostarT \Sigresstar\Gostar\right)^{-1}\GostarT\Sigstarhalf\right\}\Sigstarhalfinv\betastar_r\widehat{\Sigma}_X^{1/2}\mid
  \end{split}    
\end{equation}    
The same logic that applied to $\widehat{A}_{j,n}$ applies to $\Astar_{j,n}$. The quantity $\Sigstarhalf\Gostar\left(\GostarT \Sigresstar\Gostar\right)^{-1}\GostarT\Sigstarhalf$ in \eqref{Ajnstar} is the projection onto the column space of $\Sigstarhalf\Gostar$. The quantity $\GostarT\betastar_r \neq 0$ almost surely since $j=1,...,u-1$. As a result, the column space of $\Sigstarhalfinv\betastar_r\widehat{\Sigma}_X^{1/2}$ in \eqref{Ajnstar} has a nontrivial intersection with the column space of $\Sigstarhalf\Gostar$ almost surely. Therefore $\Astar_{j,n} > 1$ almost surely. The steps in \eqref{weightineq-2}, applied to the starred data, yields
\begin{equation} \label{weightineq-2-star}
  \surd{n}\wstar_{j} \leq  \surd{n}e^{-n\mid O_p(1)\mid}.
\end{equation}
Thus,
\begin{align*}
  \surd{n}\|\wstar_j\text{vec}(\betastar_j) - w_j\text{vec}(\hat{\beta}_j)\| 
    &\; \leq \; \surd{n}\|\wstar_j\text{vec}(\betastar_j)\| + \surd{n}\|w_j\text{vec}(\hat{\beta}_j)\| \\
    &\: \leq \: \surd{n}e^{-n\mid O_p(1)\mid}\|\text{vec}(\betastar_j)\| + \surd{n}e^{-n\mid O_p(1)\mid}\|\text{vec}(\betahat_j)\| \\
    & = 2O_p(1)\surd{n}e^{-n\mid O_p(1)\mid}
\end{align*}
for $j = 1,...,u-1$ where $\|\text{vec}(\betahat_j)\|$ and $\|\text{vec}(\betastar_j)\|$ are both $O_p(1)$ just as in the $j = u+1,...,r$ case. Combining all of these term yields the $2(u-1)O_p(1)\surd{n}e^{-n\mid O_p(1)\mid}$ order in \eqref{opmn} in \citet*{eck2}. This completes the proof when $j = 1,...,u-1$.

The final case is when $j = u$. Let $E_n = \sum_{i\neq u}^r e^{b_u - b_i}$. We can write 
$
  w_{u} = \frac{1}{1 + E_n} = 1 - \frac{E_n}{1 + E_n}.
$
The term $E_n = O_{p}\left(n^{-p}\right)$ since $e^{-n \mid O_p(1)\mid} = O_{p}\left(n^{-p}\right)$. Therefore 
\begin{align*}
  \surd{n}\wstar_u\text{vec}(\betastar_u) &= \surd{n}\left(1 - \frac{E_n}{1 + E_n}\right)\text{vec}(\betastar_u) \\
    &= \surd{n}\text{vec}(\betastar_u) + O_{p}\left\{n^{(1/2-p)}\right\}, \\
  \surd{n}w_u\text{vec}(\hat{\beta}_u) &= \surd{n}\left(1 - \frac{E_n}{1 + E_n}\right)\text{vec}(\hat{\beta}_u) \\
    &= \surd{n}\text{vec}(\hat{\beta}_u) + O_{p}\left\{n^{(1/2-p)}\right\}.
\end{align*}
Adding the previous results over $j$ to form $\surd{n}\left\{ \text{vec}(\betastar_{w}) - \text{vec}(\hat{\beta}_{w}) \right\}$ yields the result given in \eqref{opmn} in \citet*{eck2}. This completes the proof.
\end{proof}


\end{document}